\documentclass[floatfix,aps,physrev,twocolumn,unsortedaddress,showkeys,10pt]{revtex4-2}

\usepackage{graphicx}
\usepackage{siunitx}
\usepackage{physics}
\usepackage{bbold}
\usepackage[normalem]{ulem}

\begin{document}

\title{\textbf{Validating the calibrated creation of heralded single photons}}

\author{Daniel Borrero Landazabal}
\email{daniel.borrerolandazabal@dlr.de}
\author{Kaisa Laiho}
\affiliation{German Aerospace Center (DLR e.V.), Institute of Quantum Technologies, Wilhelm-Runge-Str. 10, 89081 Ulm, Germany}

\date{\today}

\begin{abstract}
Coincidence-count discrimination is a practical and straightforward tool for the characterization of photon-pair processes and heralded single photons. Here, we implement a heralded single photon source based on parametric down-conversion in a periodically poled KTiOPO$_4$ waveguide in the telecom wavelength range involving a low number of optical modes. We extend the toolbox for the loss-tolerant state characterization by combining conventional figures-of-merit in order to access the heralded state's mean photon number and its photon-number parity. Our experiment demonstrates that an accurate determination of these characteristics is possible just through simple photon-correlation measurements. We believe that our results can find usage in the calibrated creation of heralded single photons and in determining the expectation values of observables that are crucial for attributing a single quantum.
\end{abstract}


\maketitle

\section{Introduction}

Traditionally, either the homodyne detection of field quadratures \cite{lvovsky2001quantum} or the direct probing via the reconstruction of photon-number statistics of light \cite{laiho2010probing, nehra2019state} has been employed in the loss-tolerant quantum optical state characterization of free-propagating heralded single photons. Both these measurement methods deliver access to the Wigner function that offers a complete phase‑space representation of an optical quantum state. However, its experimental determination is notoriously demanding and the existing reconstruction methods require rather heavy mathematical routines. While the homodyne detection often involves a back-transformation and optimization \cite{lvovsky2009continuous, leonhardt1997measuring}, the direct probing usually includes an inversion of losses \cite{krishnaswamy2024experimental}, which is an ill-posed problem \cite{kiss1995compensation,starkov2009numerical, banaszek1999direct}, and may also involve other detector-dependent transformations related, for example, to the detector's ability to resolve photon numbers. 

Besides, there exist loss-independent state classification methods that are based on coincidence-count discrimination, like the famous Hanbury-Brown-Twiss (HBT) experiment \cite{brown1956correlation}. In the past, such experiments have been used for measuring the higher-order normalized Glauber-correlation functions of light that can be used for the state classification \cite{avenhaus2010accessing}. This method offers a more direct access to the state's characteristics with a lower experimental complexity and without the need for loss-inversion or other transformations. However, for extracting the normalized photon-correlations accurately, care has to be taken that the used single-photon sensitive detectors do not get saturated by the incoming photon flux. Additionally, without the knowledge of the mean photon number of the state, these values cannot give an access to an in-depth state characterization \cite{barnett2002methods}.

Optical energy quanta such as single photons are a cornerstone in quantum metrology. An important application area for single photons is quantum radiometry \cite{polyakov2009quantum,chunnilall2014metrology}, which strives for an accurate determination of the properties of quantized light and for its exploitation in precision measurements  \cite{ kuck2022single, georgieva2024disseminable}. In the ideal case, a single photon is emitted into a single optical mode and its photon-number distribution contains only the one-photon component \cite{eisaman2011invited,migdall2013single}. Consequently, the absence of multiphoton contributions is routinely verified by measuring the normalized second‑order correlation of the heralded state, $g_{\text{h}}^{(2)}$, in an HBT configuration. A vanishing value of $g_{\text{h}}^{(2)}$ is used as a confirmation that the emitted light exhibits a sub-Poissonian photon-number distribution \cite{sempere2022reducing,davis2022improved,magnoni2024toward,azuma2024heralded}. While this criterion rules out multiphoton emission, it does not guarantee that the mean photon number of the heralded state equals to unity. Verifying this is therefore essential for a single‑photon source. Moreover, one can go beyond by determining the photon-number parity that gives a direct access to the phase-space characteristics of light and, for example, to the single-photon non-classicality \cite{nogues2000measurement,kenfack2004negativity}.

Regarding the collection of photon pairs, typically the coincidences-to-accidentals ratio (CAR) and Klyshko's efficiencies are used for their verification \cite{klyshko1977utilization, meyer2020single}. While the former delivers the strength of producing photons in pairs, the latter describes the collection efficiency of the entire experimental arrangement taken that photons are created in pairs. Indeed, both these figures-of-merit of photon-pair production vastly govern the quality of the heralded single photons \cite{borrero2025advancing}.

Here, we implement a parametric down-conversion (PDC) process in a periodically-poled KTiOPO$_4$ (PP-KTP) waveguide in the telecom wavelengths, which produces cross-polarized photon pairs with an optical mode number $<2$. After the conventional characterization, we loss-tolerantly extract the mean photon-number and the photon-number parity of the heralded state just by counting coincidence and single counts. Our results show that both these values are utterly sensitive to the inevitable multiphoton contributions of the PDC process. We provide a boundary for the value of the CAR, above which we can reliably generate heralded single photons and approach the ultimate negative limit of the photon-number parity. Furthermore, our results show that the CAR can be employed as a calibration tool in the heralded state preparation and that one cannot solely rely on the values of $g_{\text{h}}^{(2)}$, when determining the photon-number content of the heralded single photons. 

\begin{figure}[t]
    \centering
    \includegraphics[width=1\linewidth]{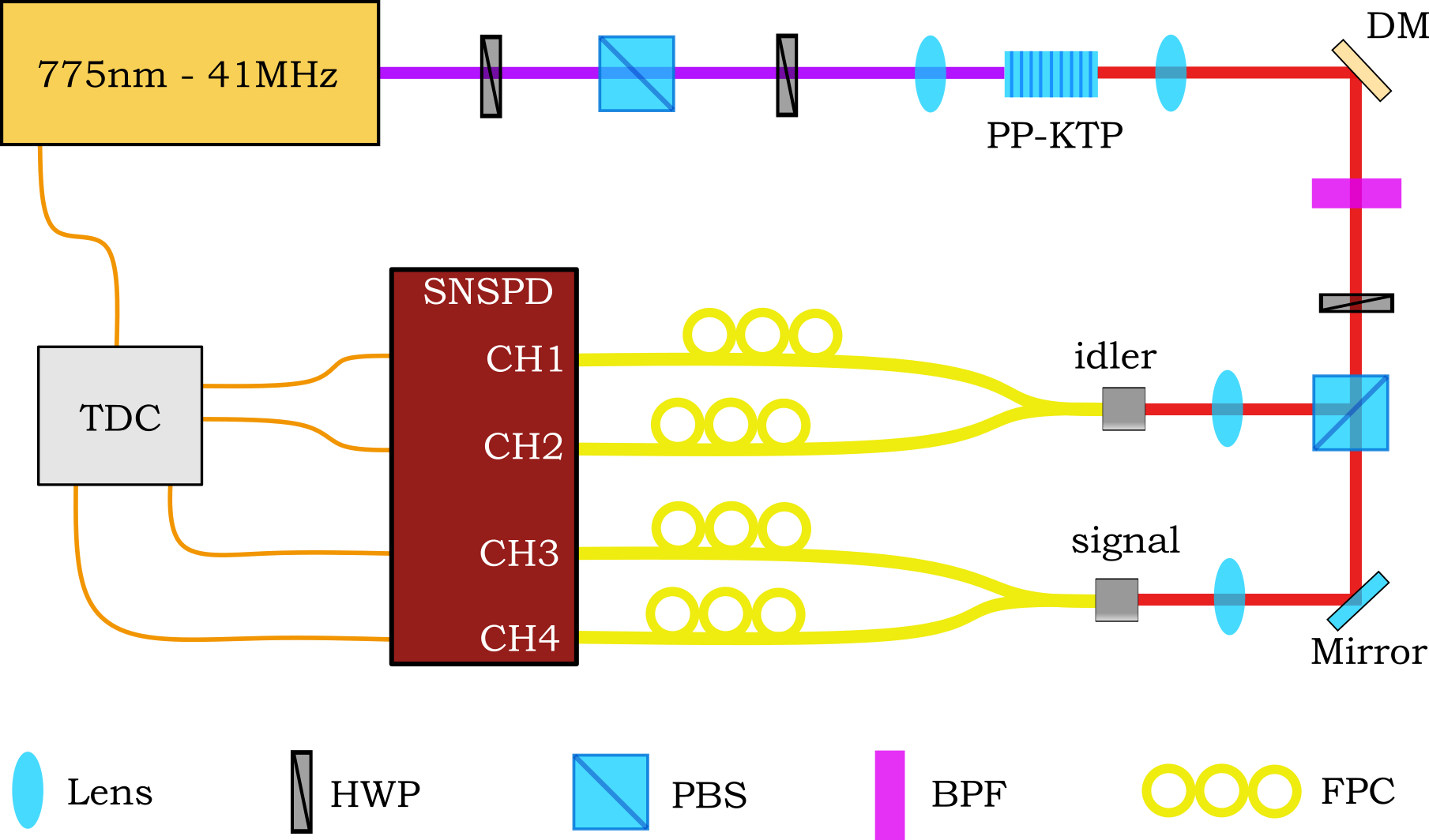}
    \caption{Experimental setup for the generation and characterization of heralded single photons. For details and abbreviations, see the main text.}
    \label{fig:setup}
\end{figure}

\section{Experimental arrangement}

Our experimental setup is sketched in Fig.~\ref{fig:setup}. The pump light centered at the wavelength of \SI{775}{\nano \meter} is produced by a pulsed laser system (Stuttgart Instruments) having the repetition rate of \SI{41}{\mega\hertz} and the pulse duration of around \SI{350}{\femto\second}. The pump light first passes through a spectral and spatial mode control (not shown) to make sure that no redundant light is on the beam path, and to guarantee a Gaussian spatial beam shape. Then, it is sent through a half-wave plate (HWP) and a polarizing beam splitter (PBS) to control the pump power. Thereafter, another HWP is used to select the proper polarization for the pump light. The pump light is then coupled into  the PP-KTP waveguide (AdvR Inc.) that has the length of \SI{11}{mm} and a cross-sectional area of around \numproduct{4x4}~\unit{\micro\meter^2} with an aspheric lens having the focal length of \SI{6.24}{mm}. Another aspheric lens with a focal length of \SI{3.1}{mm} is used for coupling the light out of the waveguide. Our waveguide produces degenerate type-II PDC such that the created signal and idler beams are cross-polarized.

The collimated PDC emission is reflected by a dichroic mirror (DM) and then passed through a bandpass filter (BPF) centered at \SI{1550}{\nano \meter} with \SI{12}{\nano \meter} bandwidth in order to separate the residual pump light from it. Thereafter, the signal and idler beams are sent through a HWP and then separated at a PBS. An optimal control of the polarization is important to minimize the leakage of the individual twin beams into the each other's beam paths. Finally, the signal and idler beams are each coupled into single-mode fibers with aspheric lenses having the focal lengths of \SI{5.5}{mm}. Both signal and idler are coupled through a fiber-optic 50:50 beam splitter and sent through fiber polarization controllers (FPCs) before connecting them into a superconducting nanowire single-photon detector (SNSPD) having four channels that are sensitive in the telecommunication wavelength range. The implemented detectors have near \SI{1550}{\nano\meter} an efficiency $>$~\SI{75}{\%}, jitter $<$~\SI{20}{\pico\second} in the full width at half maximum, and dark count rates below \SI{100}{\hertz}. The measured count rates are recorded with the help of a time-to-digital converter (TDC), triggered by the laser's photodiode. At each measured pump power we collect enough data to deduce the statistical error of the reported quantities. We employ tight time-gating by using a detection windows of $\SI{0.5}{\nano\second}$ in order to suppress the effect of dark counts.

\section{Results and discussion}

We start by extracting the conventional characteristics of the photon-pair production. For this purpose, we determine the values of CAR in terms of the pump power as well as the Klyshko's efficiencies for signal and idler. We note that CAR works for us as a calibration parameter, in terms of which we in the following present the heralded state properties. Fortunately, if all light impinging on one individual detector suffers from the same loss, the CAR represents a loss-independent figure-of-merit \cite{avenhaus2010accessing}. Therefore, it is more practical to use the CAR values for calibration than the pump power or the mean photon number of the PDC that both are highly dependent on the underlying photonic realization.

The CAR, or equivalently the first-order cross-correlation between signal-idler beams \cite{christ2011probing,allevi2012measuring,laiho2022measuring}, can be obtained in the pulsed regime via 
\begin{eqnarray}
    \text{CAR} &=& \frac{\frac{C(\text{i,s})}{\text{R}_{\text{pump}}}}{\frac{S_{\text{i}}}{\text{R}_{\text{pump}}}\times \frac{S_{\text{s}}}{\text{R}_{\text{pump}}}} \label{eq:CAR_Pump} \\
    &=& \text{R}_{\text{pump}}  \frac{C(\text{i,s})}{S_{\text{i}}\times S_{\text{s}}} = \frac{C(\text{i,s})}{C_{\text{acc}}} \nonumber \, ,
\end{eqnarray}
in which $C(\text{i,s})$ is the coincidence rate between signal and idler, $C_{\text{acc}} = S_{\text{i}}\times S_{\text{s}}/\text{R}_{\text{pump}} $ is the accidentals rate, $S_{\text{i}}$  is the single count rate of the idler, $S_{\text{s}}$ is that of signal and $\text{R}_{\text{pump}}$ is the repetition rate of the pump laser. We notice that on the right hand side of Eq.~(\ref{eq:CAR_Pump}) the nominator represents the probability of detecting a photon pair, while the terms in the denominator are the probabilities of detecting a single count in the idler or signal beam.

\begin{figure}
    \centering
    \includegraphics[width=1\linewidth]{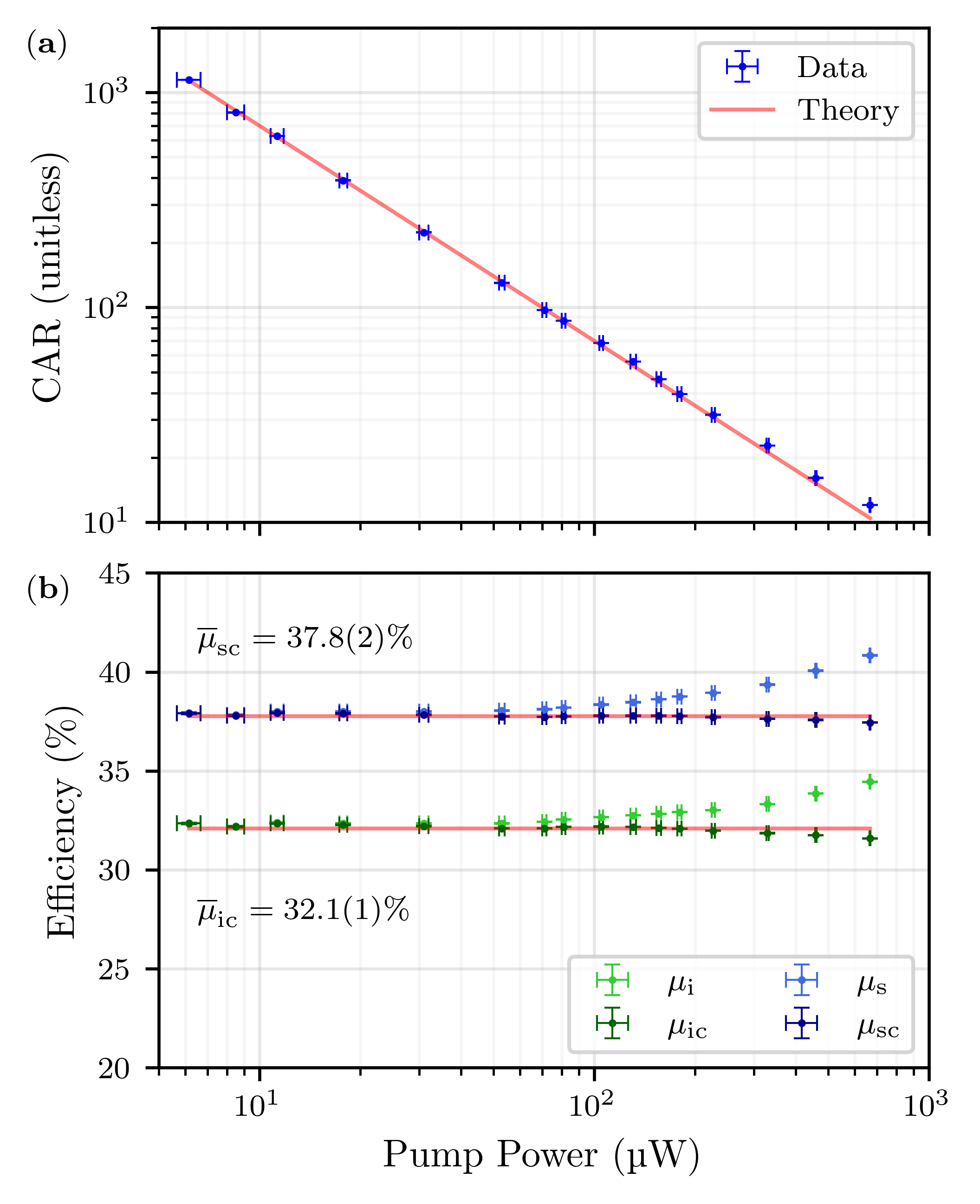}
    \caption{Conventional photon-pair characteristics in terms of pump power. (a) The measured values of the CAR are computed via Eq.~(\ref{eq:CAR_Pump}). The red solid line illustrates a theoretical fit being inversely proportional to the pump power. (b) The overall detection efficiencies in the idler (light green dots) and signal (light blue dots) beams are power dependent, whereas the corrected ones given by Eq.~(\ref{eq:klyshko}) for idler (green dots) and signal (blue dots) deliver a constant value, which is desired for a robust source. The red solid lines represent the averages of the values extracted with Eq.~(\ref{eq:klyshko}) for signal and idler.}
    \label{fig:CAR_pump}
\end{figure}

To avoid the region of high pump powers that can lead to high-gain PDC \cite{triginer2020understanding}, we restrict the CAR between $10$ to $10^3$ and illustrate the reached values in Fig.~\ref{fig:CAR_pump}(a) in terms of the pump power. As expected, the CAR follows an inverse proportionality to the pump power, allowing us to replace the pump power by the value of CAR. The overall detection efficiencies are typically determined via the Klyshko's coefficients \cite{klyshko1977utilization} and can be retrieved from coincidence counting between signal and idler via $ \mu_{\text{s/i}} = \frac{C(\text{i,s})}{S_{\text{i/s}}}$. Nevertheless, we correct the efficiencies for the accidental count rates and re-write them as \cite{krapick2014bright}
\begin{equation}
    \mu_{\text{sc/ic}} = \frac{C(\text{i,s})-C_{\text{acc}}}{S_{\text{i/s}}} \, .
\label{eq:klyshko}
\end{equation}
As presented in Fig.~\ref{fig:CAR_pump}(b), we record almost constant values for the corrected efficiencies of \SI{37.8\pm 0.1}{\%} and \SI{32.1\pm0.2}{\%} for signal and idler, respectively. 

\begin{figure}[t]
    \centering
    \includegraphics[width=1\linewidth]{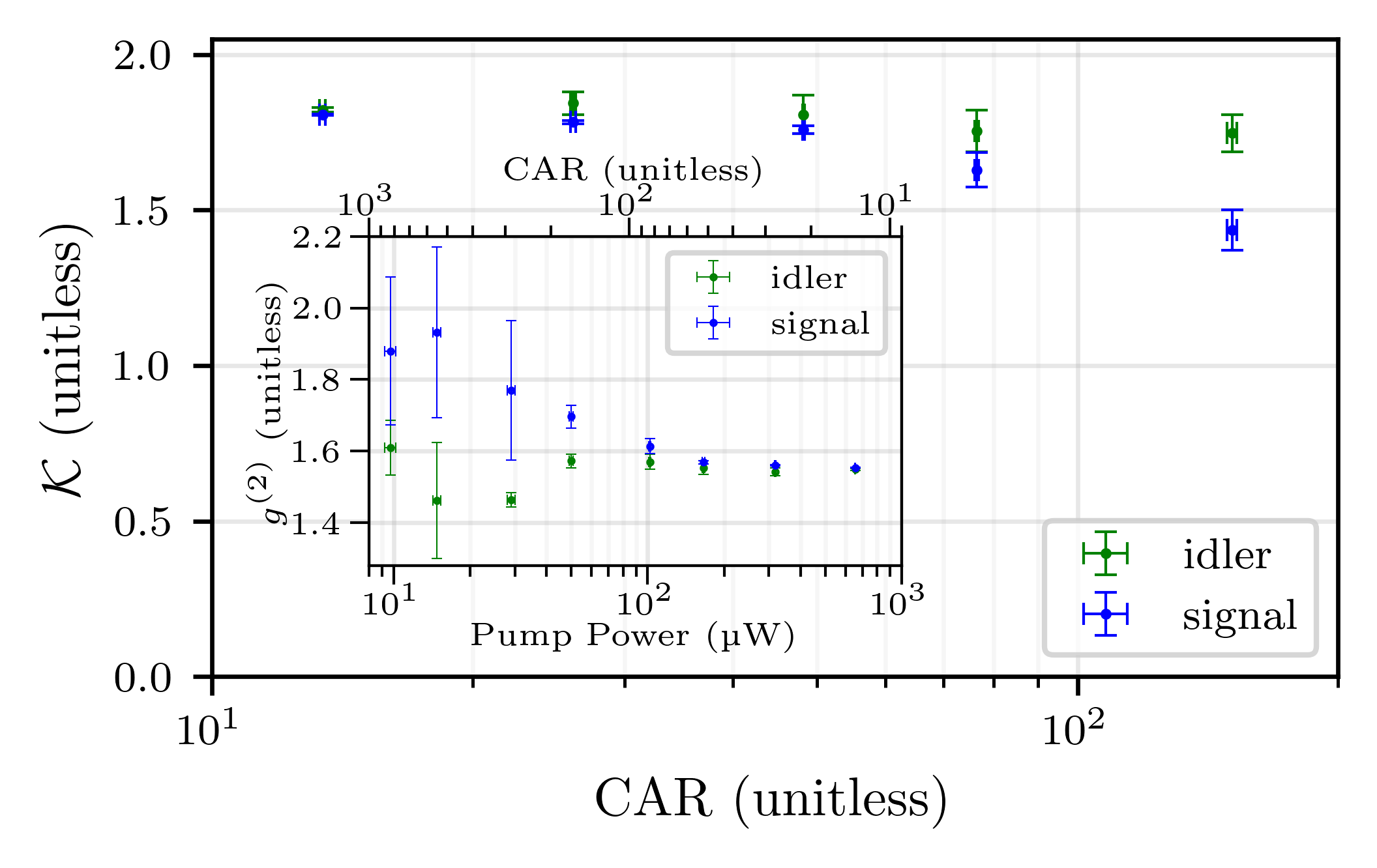}
    \caption{The extracted $\mathcal{K}$ number for signal (blue symbols) and idler (green symbols) versus the CAR. The inset shows the measured values of the unconditional $g^{(2)}$ in terms of the pump power.}
    \label{fig:Schmidt_number}
\end{figure}

Next, we use the HBT measurement to access the mode number of our PDC emission via individual marginal beams \cite{christ2011probing} that we denote as unconditional $g^{(2)}$. We expect that the unconditional $g^{(2)}$ delivers a constant value when changing the pump power \cite{laiho2011testing}, and that it ideally takes the value of two for the single-mode twin-beam PDC emission that PP-KTP platforms are known to closely produce in the telecom wavelength range \cite{evans2010bright,eckstein2011highly}. Our results are shown in the inset of Fig.~\ref{fig:Schmidt_number} in terms of the pump power. We find a value that remains rather unaltered over an order of magnitude change in the pump power. We believe that experimental imperfections such as the leakage of the other twin beam into the measured beam's path cause discrepancies at low pump powers. 

In case of PDC emission, the unconditional $g^{(2)}$ is directly related to the effective number of optical modes via \cite{christ2011probing}
\begin{equation}
    \mathcal{K} = \frac{1}{g^{(2)}-1} \, ,
     \label{eq:Schmidt_number}
\end{equation}
which is often called the Schmidt number \cite{miatto2012spatial}. We note that the $\mathcal{K}$ parameter delivers only the effective mode number assuming their uniform occupation \cite{eberly2006schmidt} and that Eq.~(\ref{eq:Schmidt_number}) applies exclusively in the region of modest pump powers. In Fig.~\ref{fig:Schmidt_number}, we illustrate the extracted mode numbers and find the average $\mathcal{K}$ number of \SI{1.67\pm0.02}{} and \SI{1.79\pm0.03}{} for signal and idler, respectively. We summarize the results for the photon pair characterization in the table~\ref{tab:PhotonPairs}.

\begin{table}[b]
\centering
\caption{Parameters characterizing the photon-pair creation process.}
\begin{tabular}{cccc}
 $ \Bar{\mu}_{\text{sc}} (\%)$ \quad\quad& $ \Bar{\mu}_{\text{ic}}(\%)$ \quad\quad& $\mathcal{K}_\text{s}$ \quad\quad& $\mathcal{K}_\text{i}$ \vspace{2mm} \\ 
\hline \vspace{-5mm}\\ \hline \\
\SI{37.8\pm 0.1}{}  \quad\quad   &  \SI{32.1\pm0.2}{}  \quad\quad   &  \SI{1.67\pm0.02}{} \quad\quad   &   \SI{1.79\pm0.03}{} 
\end{tabular}
\label{tab:PhotonPairs}
\end{table}

Next, we investigate the heralded state characteristics and start by extracting the heralded second-order correlation $(g_{\text{h}}^{(2)})$ via \cite{davis2022improved,sempere2022reducing}
\begin{eqnarray}
    g^{(2)}_{\text{h}} &=& \frac{\frac{C(\text{i},\text{s}_1,\text{s}_2)}{S_{\text{i}}}}{\frac{C(\text{i},\text{s}_1)}{S_{\text{i}}}\times\frac{C(\text{i},\text{s}_2)}{S_{\text{i}}}}  \label{eq:g2h} \\
    &=& S_{\text{i}}\frac{C(\text{i},\text{s}_1,\text{s}_2)}{C(\text{i},\text{s}_1)\times C(\text{i},\text{s}_2)} \nonumber \, ,
\end{eqnarray}
in which $C(\text{i},\text{s}_1,\text{s}_2)$ is the rate of the heralded coincidences in signal and $C(\text{i},\text{s}_{1(2)})$ are the coincidence rates between herald and individual signal channels, while $S_{\text{i}}$ is the singles rate in idler now representing the heralding rate. Similar to Eq. (\ref{eq:CAR_Pump}) the ratios on the right-hand side of Eq.~(\ref{eq:g2h}) now correspond to the coincidence and single click probabilities in the signal arm conditioned on a click in the herald.

\begin{figure}[t]
    \centering
    \includegraphics[width=1\linewidth]{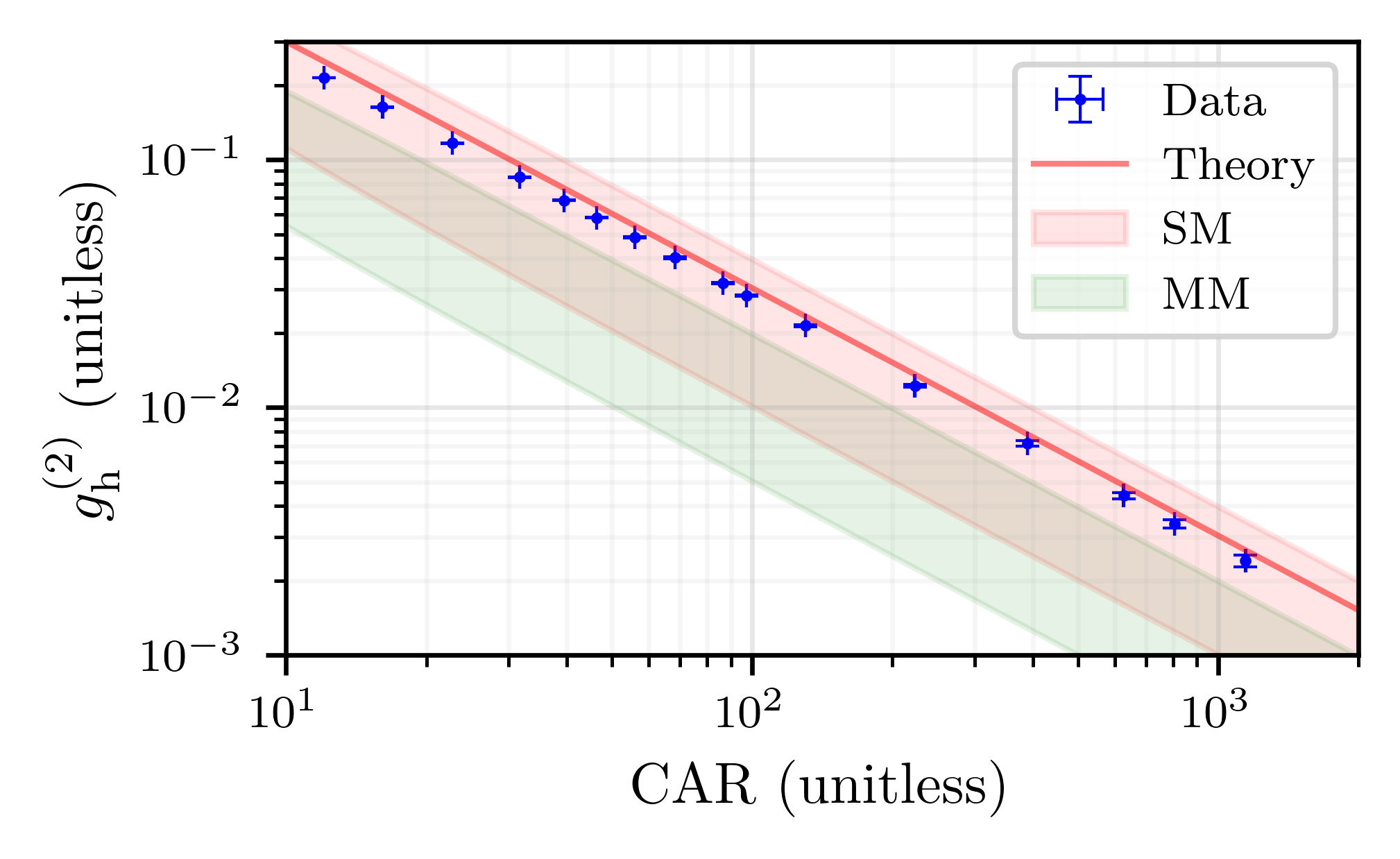}
    \caption{Heralded second-order correlation function in terms of CAR. The red(green)-shaded area corresponds to values with different heralding efficiency for a single(multi)-mode twin-beam PDC emission.}
    \label{fig:g2h}
\end{figure}

We illustrate the results for $g_{\text{h}}^{(2)}$ in Fig.~\ref{fig:g2h} in terms of the CAR values. The symbols represent the measured data, whereas the theory (red line), which is calculated following the treatment in Ref.~\cite{borrero2025advancing} that applies only in the low gain regime (CAR $\gtrsim 5-10$) \cite{christ2011probing}, corresponds to the case of a single-mode twin-beam state having the same heralding efficiency of $\SI{32.1}{\%}$ as in the experiment. The measured values of $g_{\text{h}}^{(2)}$ agree well with the single-mode theory. Additionally, the red-shaded area depicts the possible values of $g_{\text{h}}^{(2)}$ for the single-mode twin-beam state, in the range of the heralding efficiency from $\SI{1}{\%}$ to $\SI{100}{\%}$. We emphasize that a larger heralding efficiency lowers the values of  $g_{\text{h}}^{(2)}$. The green-shaded area shows the possible values for $g_{\text{h}}^{(2)}$ in case of multimode twin-beam PDC emission \cite{multimodePDC}. We emphasize that despite the slight discrepancy in the mode-number the single-mode theory predict values close to the measured ones. We achieve for CAR $=$ \SI{97.14\pm 0.04} a value of $g_{\text{h}}^{(2)}=$ \SI{2.84\pm0.02d-2}{}. 

Finally, we investigate the mean photon number of the heralded states and thereafter extract their photon-number parity. For this purpose, we model the detection system with a positive operator-valued measure (POVM) acting on the heralded signal beam as \cite{borrero2025advancing,sperling2014quantum}
\begin{eqnarray}
    \hat{O}_k &=&  \sum_{m=0}^{k} \binom{N}{k}\binom{k}{m} (-1)^m e^{-\frac{\nu}{N}(N+m-k)} \nonumber\\
    & &\times \sum^{\infty}_{n=0}\left(1-\frac{\mu_{\text{sc}}}{N}(N+m-k)\right)^{n} \ket{n}_{\text{s~s}}\bra{n} \, ,
\label{eq:POVM}
\end{eqnarray}
where $k$ is the number of click-detections, $N$ the total amount of detectors in the signal path, $\nu$ the dark count probability and $\mu_{\text{sc}}$ the corrected efficiency of the heralded state detection. Here, we focus on the detection of single photons $(k=1)$, implement two detectors in the signal $(N=2)$, and employ the efficiency $\mu_{\text{sc}}$ from Fig.~\ref{fig:CAR_pump}. Furthermore, the dark count probability is neglected, since the tight time gating drops the dark count probability to $\sim \SI{e-8}{}$ per pulse. 

To extract the heralded-state mean photon-number, we first reduce the POVM given in Eq.~(\ref{eq:POVM}) by using the given detector parameters. Thus, the POVM is simplified to
\begin{equation}
    \hat{O}_{1} =  2 \sum^{\infty}_{n=0}\left\{\left(1-\frac{\mu_{\text{sc}}}{2}\right)^{n} - \left(1-\mu_{\text{sc}}\right)^{n}\right\}\ket{n}_{\text{s~s}}\bra{n} \, .
\label{eq:POVM_reduced}
\end{equation}
Furthermore, we expand the reduced POVM with a Taylor expansion and add similar terms up,
\begin{equation}
    \hat{O}_{1} =  \sum^{\infty}_{n=0}\left\{\mu_{\text{sc}}n-\frac{3}{4} \left(\mu_{\text{sc}}\right)^2n(n-1)\,...\right\}\ket{n}_{\text{s~s}}\bra{n} \, .
\label{eq:POVM_taylor}
\end{equation}

We then compute the expectation value of the expanded POVM over the heralded signal beam by considering its density matrix in the photon-number basis in the form
\begin{equation}
    \hat{\rho}_{\text{s}} = \sum_{n,m}\alpha_{n,m} \ket{n}_{\text{s}\hspace{1.5mm}\text{s}}\bra{m} \, ,
\label{eq:twb_state}
\end{equation}
in which the elements $\alpha_{n,m}$ depend on the PDC process parameters \cite{allevi2022multi}. Nevertheless, we do not need the knowledge of their exact form. Now, by taking the trace over signal we arrive at 
\begin{equation}
    \Tr_\text{s}\{\hat{\rho}_{\text{s}}\hat{O}_{1}\} =  \mu_{\text{sc}} \expval{\hat{n}}_\text{s} - \frac{3}{4}\mu_{\text{sc}}^2~g^{(2)}_{\text{h}}\expval{\hat{n}}_\text{s}^2 + \mathbb{O}(3) \, ,
\label{eq:Tr_s}
\end{equation}
where we made use of the relation $\expval{\hat{n}(\hat{n}-\mathbb{1})}_\text{s} = g^{(2)}_{\text{h}}\expval{\hat{n}}_\text{s}^2$ with $\mathbb{1}$ being the identity operator \cite{loudon2000quantum}. In Eq.~(\ref{eq:Tr_s}), we neglect the higher-order terms denoted with $\mathbb{O}(3)$. Nevertheless, we expect the second-order approximation to be valid for single photons regardless of the efficiency, since by measuring $g^{(2)}_\text{h} \ll 1$, it is reasonable to assume that contributions from all photon numbers $\geq 2$ are low. 

\begin{figure*}
    \centering
    \includegraphics[width=1\linewidth]{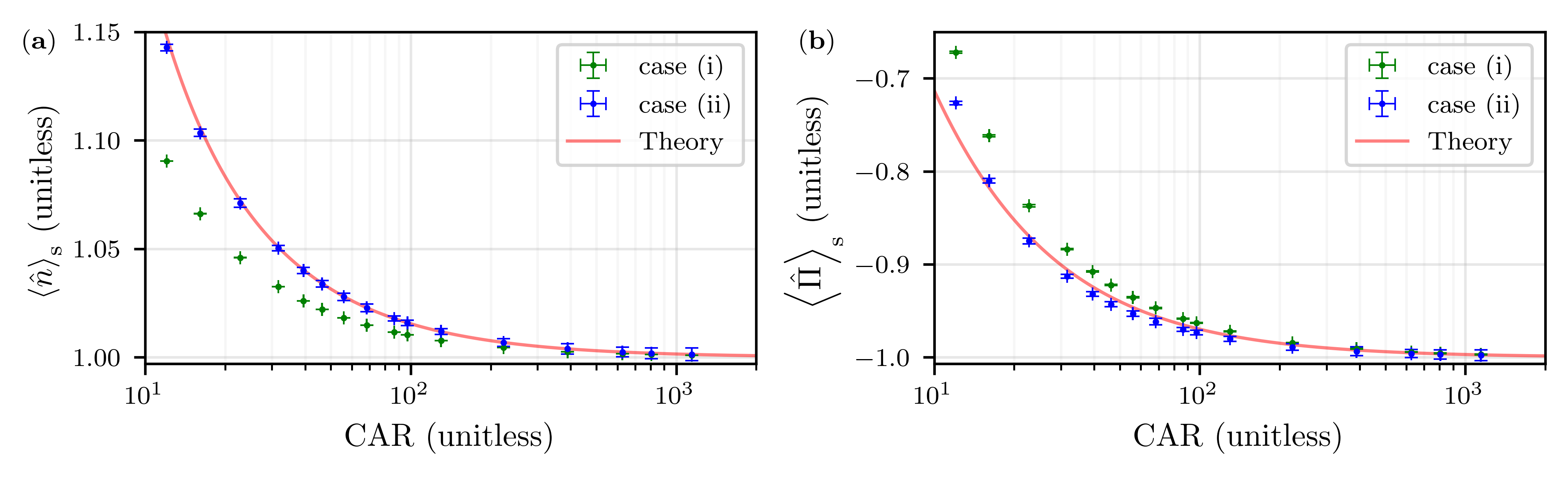}
    \caption{Expectation values for the studied observables in terms of CAR in cases (i) and (ii). When CAR increases, both (a) the mean photon number and (b) the photon-number parity approach the desired values of 1 and -1, respectively. The theoretical prediction (red line) agrees well with the second-order approximation investigated in case (ii).}
    \label{fig:mean_parity}
\end{figure*}

The trace $\Tr_\text{s}\{\hat{\rho}_{\text{s}}\hat{O}_{1}\}$ in Eq.~(\ref{eq:Tr_s}) represents the probability of detecting a heralded click in the signal arm. Experimentally, this corresponds to the probability of detecting a click in signal arm conditioned on a detection event in the herald that can be expressed via
\begin{equation}
     \Tr_\text{s}\{\hat{\rho}_{\text{s}}\hat{O}_{1}\} = \frac{C(\text{i,s})}{S_\text{i}} \equiv \mu_{\text{s}} \, ,
\end{equation}
which by definition corresponds to the Klyshko's coefficient of the signal beam.
Thus, the mean photon number of the heralded single photon can be extracted by solving Eq.~(\ref{eq:Tr_s}) for $\expval{\hat{n}}_\text{s}$, thus allowing its evaluation from photon-correlations measurements. In the following we extract it by considering both (i) the first-order and (ii) the second-order approximation of Eq.~(\ref{eq:Tr_s}). It is expected that in the postselected ensemble of the heralded-state statistics the mean photon number approaches unity. Therefore, in contrast to the prior measurements, the click-detectors can reach a saturation \cite{hadfield2009single}. Nevertheless, we expect the correction from the second-order approximation to be small.

In case (i), which applies at small detection efficiencies $\mu^2_{\text{sc}}<<1$, we retrieve \cite{laiho2019photon}
\begin{equation}
    n^{\text{1st}} =  \frac{\mu_\text{s}}{ \mu_{\text{sc}}} = \frac{\text{CAR}}{\text{CAR} - 1} \, .
\label{eq:firstorder}
\end{equation}

However, at high detection efficiencies as in our case (table~\ref{tab:PhotonPairs}) the condition for the first order approximation is not satisfied. The derivation in case (ii) delivers 
\begin{equation}
    n^{\text{2nd}} = \frac{1 - \sqrt{1-3 g^{(2)}_\text{h} \mu_{\text{s}}}}{\frac{3}{2}\mu_{\text{sc}}~ g^{(2)}_\text{h}} \, ,
\end{equation}
which not only depends on the efficiencies, but also on the heralded $g^{(2)}_\text{h}$. We further note that we only regard one of the two solutions of the quadratic formula in Eq.~(\ref{eq:Tr_s}) since only one of them delivers physically congruent results and leads to Eq.~(\ref{eq:firstorder}) in the limit $ g^{(2)}_\text{h}\rightarrow 0$.

We present the deduced values for the mean photon number of the heralded single photons in Fig.~\ref{fig:mean_parity}(a). As expected, the values extracted in case (i) presented with green symbols deviate much at low values of CAR from the theoretical expectation (red solid line), while the results of case (ii) agree well with that. Here, the theory represents an exact calculation and is obtained following the treatment in Ref.~\cite{borrero2025advancing}. The agreement between case (ii) and the theory confirms that the second-order expansion is sufficient at all values of CAR for covering the effect of multiphoton contamination. We make further emphasize that both cases (i) and (ii) converge to the same result at high CAR values. This implies that the first-order approximation is completely adequate in the region CAR $>100$. Additionally, the mean photon number approaches unity asymptotically for increasing values of CAR, which is the case for negligible dark count probability \cite{borrero2025advancing}. For instance, in case (ii) we reach at the value of CAR $=$ \SI{97.14\pm 0.04} the mean photon number $\expval{\hat{n}}_\text{s}=$ \SI{1.016\pm0.001}{}.

Moreover, we investigate the photon-number parity of the heralded single photon state \cite{laiho2019photon}. This quantity can be extracted from the mean photon number and $g^{(2)}_\text{h}$ via \cite{barnett2002methods,cahill1969density} 
\begin{equation}
    \expval{\hat{\Pi}}_\text{s} = \sum_m g^{(m)}_\text{h} \frac{(-2\expval{\hat{n}}_\text{s})^m}{m!} \approx 1 - 2\expval{\hat{n}}_\text{s} + 2\expval{\hat{n}}^2_\text{s} g^{(2)}_\text{h} \, ,
\end{equation}
on the right-hand side of which we have expanded the summation up to the second order. 

The results of the photon-number parity for the heralded single photon state are shown in Fig.~\ref{fig:mean_parity}(b). Again, we present the results for the case (i) and case (ii) with green and blue symbols, respectively, together with the theoretical prediction (red line). Once more, our results show the necessity of the second-order approximation at low values of CAR ($<\SI{e2}{}$) and that the expectation value of the photon-number parity strongly deviates from the value $\SI{-1}{}$, which is expected for ideal single photons being a further indication of multiphoton contamination in the heralded state. In case (ii), at the value of CAR $=$ \SI{97.14\pm 0.04} we achieve $\expval{\hat{\Pi}}_\text{s}=$\SI{-0.973\pm 0.002}{}. Similarly, in the region above that, the first- and second-order approximations hardly deviate from each other, and the expectation value of the photon-number parity approaches the ideal negative limit.

\begin{table*}
\centering
\begin{tabular}{ccccc} 
Work & Source & Detectors & \hspace{2mm}  Efficiency & Measured properties
\vspace{2mm} \\ 
\hline \vspace{-5mm}\\ \hline \vspace{-2mm}\\ 

This work & PP-KTP & HBT & $32.1(2)\%$& \hspace{3mm} $g^{(2)}_{\text{h}} = 0.0024(1)$, $\expval{\hat{n}}_{\text{s}}=1.001(3)$, $\expval{\hat{\Pi}}_{\text{s}}=-0.998(6)$ \\ \\

\cite{laiho2010probing} & KTP & TMD & $16.5\%$ & $\expval{\hat{\Pi}}_{\text{s}}=-0.888(6) ^\dagger$ \\ \\

\cite{nehra2019state} & PP-KTP & PNR & $58\%$ &  $\expval{\hat{\Pi}}_{\text{s}}=-0.055(8) ^{\dagger, \star} $  \\ \\

\cite{laiho2019photon} & AlGaAs BRW & PNR & $ 4\%$ & $g^{(2)}_{\text{h}} = 0.21(3)$, $\expval{\hat{n}}_{\text{s}}=1.086(2)$, 
$\expval{\hat{\Pi}}_{\text{s}} =-0.68(8)$ \\ \\

\cite{harder2016local} & PP-KTP & TMD & $15\%^\diamond$ &  $\expval{\hat{\Pi}}_{\text{s}}>-0.3$ \\ \\

\cite{lam2025optimizing} & Quantum Dot & HBT & $3\%$ &  $g^{(2)}_{\text{h}} = 0.0412(5)$ 

\end{tabular}
\caption{Comparison of our work with different direct probing experiments for single-photon states. PNR: photon-number resolving, TMD: time-multiplexed detection, and BRW: Bragg-reflection waveguide. $^\dagger$The photon-number parity was extracted from the reported value of the Wigner function at the origin. $^\star$Value includes no loss-inversion. $^\diamond$Value is estimated as a product of the reported detection and coupling efficiencies.}
\label{tab:Comparison}
\end{table*}

Lastly, in table~\ref{tab:Comparison} we compare different measurements aiming at direct probing of the single-photon characteristics \cite{banaszek1996direct,wallentowitz1996unbalanced}. Our realization is resource saving as it only uses manifold coincidence discrimination and works, for example, without true photon-number resolving detectors or sophisticated mathematics related to the reconstruction expectation values from click statistics. Further, we achieve precision in measuring the studied expectation values loss-tolerantly and can set calibration boundaries for the underlying photon-pair generation process for the heralded-state production.

\section{Conclusions.}

Heralded single photons are an important resource in quantum optics experiments, and fast and easy tools are required for their effective characterization. For this purpose, photon correlation measurements provide a simple and adequate toolbox. We generate heralded single photons from a PP-KTP waveguide in the telecom wavelength range with an optical mode number of \SI{1.67\pm0.02}{} and \SI{1.79\pm0.03}{} for signal and idler, respectively. Further, we take use of the loss-independent values of the CAR as a calibration parameter with respect to which the studied figures-of-merit are presented. After conventional characterization of the source, we loss-tolerantly determine the mean photon number and the photon-number parity of heralded single photons just by singles and coincidence counting. Our results show their asymptotic behavior and that the ideally expected values are approached when CAR is strongly increased, indicating a drastic reduction of the multiphoton contributions. We showed that at a high heralding efficiency of \SI{32.1\pm0.2}{\%} a second-order approximation is necessary in order to accurately extract these parameters. In the region of CAR $=$ \SI{97.14\pm 0.04} we achieved a value of $g_{\text{h}}^{(2)}=$ \SI{2.84\pm0.02d-2}{} and loss-tolerantly extracted for the mean photon number $\expval{\hat{n}}_\text{s}=$ \SI{1.016\pm0.001}{} and photon-number parity $\expval{\hat{\Pi}}_\text{s}=$\SI{-0.973\pm 0.002}{}. Our measurement is easy and fast to implement in comparison to other available methods such as the homodyne tomography or the photon-statistics reconstruction. We are confident that the utilized measurement tool represents an appropriate method for accurately accessing the heralded state's properties, ease up the comparison of heralded single photon sources and enable their calibrated usage.

\section{Acknowledgements}
We thank Thomas Dirmeier and Christoph Marquardt for support with the laboratory equipment and Freyja Ullinger and Matthias Zimmermann for fruitful discussions about the photon-number parity.

\bibliography{references.bib}

@PREAMBLE{
 "\providecommand{\noopsort}[1]{}" 
 # "\providecommand{\singleletter}[1]{#1}%" 
}

@book{loudon2000quantum,
  title={The quantum theory of light},
  author={Loudon, Rodney},
  year={2000},
  publisher={Oxford University Press, Oxford}
}

@book{barnett2002methods,
  title={Methods in theoretical quantum optics},
  author={Barnett, Stephen and Radmore, Paul M},
  year={1997},
  publisher={Oxford University Press, Oxford}
}

@book{migdall2013single,
  title={Single-photon generation and detection: physics and applications},
  editor={Migdall, Alan and Polyakov, Sergey V and Fan, Jingyun and Bienfang, Joshua C},
  volume={45},
  year={2013},
  publisher={Academic Press, Amsterdam}
}

@article{brown1956correlation,
  title={Correlation between photons in two coherent beams of light},
  author={Brown, R Hanbury and Twiss, Richard Q},
  journal={Nature},
  volume={177},
  number={4497},
  pages={27--29},
  year={1956},
  publisher={Nature Publishing Group UK London}
}

@article{cahill1969density,
  title={Density operators and quasiprobability distributions},
  author={Cahill, Kevin E and Glauber, Roy J},
  journal={Phys. Rev.},
  volume={177},
  number={5},
  pages={1882},
  year={1969}
}

@article{banaszek1996direct,
  title={Direct probing of quantum phase space by photon counting},
  author={Banaszek, Konrad and W{\'o}dkiewicz, Kryzysztof},
  journal={Physical review letters},
  volume={76},
  number={23},
  pages={4344},
  year={1996},
  publisher={APS}
}

@article{wallentowitz1996unbalanced,
  title={Unbalanced homodyning for quantum state measurements},
  author={Wallentowitz, S and Vogel, W},
  journal={Physical Review A},
  volume={53},
  number={6},
  pages={4528},
  year={1996},
  publisher={APS}
}

@book{leonhardt1997measuring,
  title={Measuring the quantum state of light},
  author={Leonhardt, Ulf},
  year={1997},
  publisher={Cambridge University Press, Cambridge}
}

@article{banaszek1999direct,
  title={Direct measurement of the Wigner function by photon counting},
  author={Banaszek, K and Radzewicz, C and W{\'o}dkiewicz, K and Krasi{\'n}ski, JS},
  journal={Phys. Rev. A},
  volume={60},
  number={1},
  pages={674},
  year={1999},
  publisher={APS}
}

@article{lvovsky2001quantum,
  title={Quantum state reconstruction of the single-photon Fock state},
  author={Lvovsky, Alexander I and Hansen, Hauke and Aichele, T and Benson, O and Mlynek, J and Schiller, S},
  journal={Phys. Rev. Lett.},
  volume={87},
  number={5},
  pages={050402},
  year={2001},
  publisher={APS}
}

@article{lvovsky2009continuous,
  title={Continuous-variable optical quantum-state tomography},
  author={Lvovsky, Alexander I and Raymer, Michael G},
  journal={Rev. Mod. Phys.},
  volume={81},
  number={1},
  pages={299--332},
  year={2009},
  publisher={APS}
}

@article{nehra2019state,
  title={State-independent quantum state tomography by photon-number-resolving measurements},
  author={Nehra, Rajveer and Win, Aye and Eaton, Miller and Shahrokhshahi, Reihaneh and Sridhar, Niranjan and Gerrits, Thomas and Lita, Adriana and Nam, Sae Woo and Pfister, Olivier},
  journal={Optica},
  volume={6},
  number={10},
  pages={1356--1360},
  year={2019},
  publisher={Optical Society of America}
}

@article{harder2016local,
  title={Local sampling of the Wigner function at telecom wavelength with loss-tolerant detection of photon statistics},
  author={Harder, G and Silberhorn, Ch and Rehacek, J and Hradil, Z and Motka, L and Stoklasa, B and S{\'a}nchez-Soto, Luis Lorenzo},
  journal={Physical review letters},
  volume={116},
  number={13},
  pages={133601},
  year={2016},
  publisher={APS}
}

@article{nogues2000measurement,
  title={Measurement of a negative value for the Wigner function of radiation},
  author={Nogues, G and Rauschenbeutel, A and Osnaghi, S and Bertet, P and Brune, M and Raimond, JM and Haroche, S and Lutterbach, LG and Davidovich, L},
  journal={Phys. Rev. A},
  volume={62},
  number={5},
  pages={054101},
  year={2000},
  publisher={APS}
}

@article{kenfack2004negativity,
  title={Negativity of the Wigner function as an indicator of non-classicality},
  author={Kenfack, Anatole and {\.Z}yczkowski, Karol},
  journal={J. Opt. B: Quantum Semiclass. Opt.},
  volume={6},
  number={10},
  pages={396},
  year={2004},
  publisher={IOP Publishing}
}

@article{chunnilall2014metrology,
  title={Metrology of single-photon sources and detectors: a review},
  author={Chunnilall, Christopher J and Degiovanni, Ivo Pietro and K{\"u}ck, Stefan and M{\"u}ller, Ingmar and Sinclair, Alastair G},
  journal={Opt. Eng.},
  volume={53},
  number={8},
  pages={081910--081910},
  year={2014},
  publisher={Society of Photo-Optical Instrumentation Engineers}
}

@article{polyakov2009quantum,
  title={Quantum radiometry},
  author={Polyakov, Sergey V and Migdall, Alan L},
  journal={J. Mod. Opt.},
  volume={56},
  number={9},
  pages={1045--1052},
  year={2009},
  publisher={Taylor \& Francis}
}

@article{kuck2022single,
  title={Single photon sources for quantum radiometry: a brief review about the current state-of-the-art},
  author={K{\"u}ck, Stefan and L{\'o}pez, Marco and Hofer, Helmuth and Georgieva, Hristina and Christinck, Justus and Rodiek, Beatrice and Porrovecchio, Geiland and {\v{S}}mid, Marek and G{\"o}tzinger, Stephan and Becher, Christoph and others},
  journal={Appl. Phys. B},
  volume={128},
  number={2},
  pages={28},
  year={2022},
  publisher={Springer}
}

@article{georgieva2024disseminable,
  title={Disseminable single-photon source for quantum radiometry},
  author={Georgieva, Hristina and Gerrits, Thomas and Ma, Lijun and Dawkins, Riley and L{\'o}pez, Marco and Slattery, Oliver and Kanold, Niklas and Kaganskiy, Arsenty and Rodt, Sven and Reitzenstein, Stephan and others},
  journal={Appl. Phys. Lett.},
  volume={125},
  number={26},
  year={2024},
  publisher={AIP Publishing}
}

@article{eisaman2011invited,
  title={Invited review article: Single-photon sources and detectors},
  author={Eisaman, Matthew D and Fan, Jingyun and Migdall, Alan and Polyakov, Sergey V},
  journal={Rev. Sci. Inst.},
  volume={82},
  number={7},
  year={2011},
  publisher={AIP Publishing}
}

@article{meyer2020single,
  title={Single-photon sources: Approaching the ideal through multiplexing},
  author={Meyer-Scott, Evan and Silberhorn, Christine and Migdall, Alan},
  journal={Rev. Sci. Inst.},
  volume={91},
  number={4},
  year={2020},
  publisher={AIP Publishing}
}

@article{hadfield2009single,
  title={Single-photon detectors for optical quantum information applications},
  author={Hadfield, Robert H},
  journal={Nature photonics},
  volume={3},
  number={12},
  pages={696--705},
  year={2009},
  publisher={Nature Publishing Group}
}

@article{magnoni2024toward,
  title={Toward an optical-fiber-based temporally multiplexed single-photon source},
  author={Magnoni, Agustina G and Knoll, Laura T and W{\"o}lcken, Lina and Defant, Juli{\'a}n and Morales, Juli{\'a}n and Larotonda, Miguel A},
  journal={Phys. Rev. A},
  volume={110},
  number={3},
  pages={033712},
  year={2024},
  publisher={APS}
}

@article{azuma2024heralded,
  title={Heralded single-photon source based on superpositions of squeezed states},
  author={Azuma, Hiroo and Munro, William J and Nemoto, Kae},
  journal={Phys. Rev. A},
  volume={109},
  number={5},
  pages={053711},
  year={2024},
  publisher={APS}
}

@article{sempere2022reducing,
  title={Reducing g (2)(0) of a parametric down-conversion source via photon-number resolution with superconducting nanowire detectors},
  author={Sempere-Llagostera, S and Thekkadath, GS and Patel, RB and Kolthammer, WS and Walmsley, IA},
  journal={Opt. Express},
  volume={30},
  number={2},
  pages={3138--3147},
  year={2022},
  publisher={Optica Publishing Group}
}

@article{davis2022improved,
  title={Improved heralded single-photon source with a photon-number-resolving superconducting nanowire detector},
  author={Davis, Samantha I and Mueller, Andrew and Valivarthi, Raju and Lauk, Nikolai and Narvaez, Lautaro and Korzh, Boris and Beyer, Andrew D and Cerri, Olmo and Colangelo, Marco and Berggren, Karl K and others},
  journal={Phys. Rev. Appl.},
  volume={18},
  number={6},
  pages={064007},
  year={2022},
  publisher={APS}
}

@article{laiho2022measuring,
  title={Measuring higher-order photon correlations of faint quantum light: a short review},
  author={Laiho, K and Dirmeier, T and Schmidt, M and Reitzenstein, S and Marquardt, C},
  journal={Phys. Lett. A},
  volume={435},
  pages={128059},
  year={2022},
  publisher={Elsevier}
}

@article{allevi2012measuring,
  title={Measuring high-order photon-number correlations in experiments with multimode pulsed quantum states},
  author={Allevi, Alessia and Olivares, Stefano and Bondani, Maria},
  journal={Phys. Rev. A},
  volume={85},
  number={6},
  pages={063835},
  year={2012},
  publisher={APS}
}

@article{borrero2025advancing,
  title={Advancing the heralded photon-number-state characterization by understanding the interplay of experimental settings},
  author={Borrero Landazabal, Daniel and Laiho, Kaisa},
  journal={New J. Phys.},
  volume={27},
  number={6},
  pages={064103},
  year={2025},
  publisher={IOP Publishing}
}

@article{laiho2011testing,
  title={Testing spectral filters as Gaussian quantum optical channels},
  author={Laiho, Kaisa and Christ, Andreas and Cassemiro, Kati{\'u}scia N and Silberhorn, Christine},
  journal={Opt. Lett.},
  volume={36},
  number={8},
  pages={1476--1478},
  year={2011},
  publisher={Optical Society of America}
}

@article{laiho2010probing,
  title={Probing the negative Wigner function of a pulsed single photon point by point},
  author={Laiho, Kaisa and Cassemiro, Kati{\'u}scia N and Gross, David and Silberhorn, Christine},
  journal={Phys. Rev. Lett.},
  volume={105},
  number={25},
  pages={253603},
  year={2010},
  publisher={APS}
}

@article{avenhaus2010accessing,
  title={Accessing higher order correlations in quantum optical states by time multiplexing},
  author={Avenhaus, M and Laiho, K and Chekhova, MV and Silberhorn, Ch},
  journal={Phys. Rev. Lett.},
  volume={104},
  number={6},
  pages={063602},
  year={2010},
  publisher={APS}
}

@article{laiho2019photon,
  title={Photon-number parity of heralded single photons from a Bragg-reflection waveguide reconstructed loss-tolerantly via moment generating function},
  author={Laiho, K and Schmidt, M and Suchomel, H and Kamp, M and H{\"o}fling, S and Schneider, C and Beyer, J and Weihs, G and Reitzenstein, Stephan},
  journal={New J. Phys.},
  volume={21},
  number={10},
  pages={103025},
  year={2019},
  publisher={IOP Publishing}
}

@article{christ2011probing,
  title={Probing multimode squeezing with correlation functions},
  author={Christ, Andreas and Laiho, Kaisa and Eckstein, Andreas and Cassemiro, Kati{\'u}scia N and Silberhorn, Christine},
  journal={New J. Phys.},
  volume={13},
  number={3},
  pages={033027},
  year={2011},
  publisher={IOP Publishing}
}

@article{allevi2022multi,
  title={Multi-mode twin-beam states in the mesoscopic intensity domain},
  author={Allevi, Alessia and Bondani, Maria},
  journal={Phys. Lett. A},
  volume={423},
  pages={127828},
  year={2022},
  publisher={Elsevier}
}

@article{lam2025optimizing,
  title={Optimizing the quantum interference between single photons and local oscillator with photon correlations},
  author={Lam, Hubert and {\'A}lvarez, Juan R and Steindl, Petr and Maillette de Buy Wenniger, Ilse and Wein, Stephen and Pishchagin, Anton and Huong Au, Thi and Boissier, Sebastien and Lema{\^\i}tre, Aristide and L{\"o}ffler, Wolfgang and others},
  journal={Quantum Science and Technology},
  volume={10},
  number={4},
  pages={045061},
  year={2025},
  publisher={IOP Publishing}
}

@article{triginer2020understanding,
  title={Understanding high-gain twin-beam sources using cascaded stimulated emission},
  author={Triginer, Gil and Vidrighin, Mihai D and Quesada, Nicol{\'a}s and Eckstein, Andreas and Moore, Merritt and Kolthammer, W Steven and Sipe, JE and Walmsley, Ian A},
  journal={Physical Review X},
  volume={10},
  number={3},
  pages={031063},
  year={2020},
  publisher={APS}
}

@article{evans2010bright,
  title={Bright Source of Spectrally Uncorrelated Polarization-Entangled Photons with Nearly Single-Mode Emission},
  author={Evans, Philip G and Bennink, Ryan S and Grice, Warren P and Humble, Travis S and Schaake, Jason},
  journal={Physical Review Letters},
  volume={105},
  number={25},
  pages={253601},
  year={2010},
  publisher={APS}
}

@article{eckstein2011highly,
  title={Highly efficient single-pass source of pulsed single-mode twin beams of light},
  author={Eckstein, Andreas and Christ, Andreas and Mosley, Peter J and Silberhorn, Christine},
  journal={Physical Review Letters},
  volume={106},
  number={1},
  pages={013603},
  year={2011},
  publisher={APS}
}

@article{starkov2009numerical,
  title={Numerical reconstruction of photon-number statistics from photocounting statistics: Regularization of an ill-posed problem},
  author={Starkov, VN and Semenov, AA and Gomonay, HV},
  journal={Phys. Rev. A},
  volume={80},
  number={1},
  pages={013813},
  year={2009},
  publisher={APS}
}

@article{kiss1995compensation,
  title={Compensation of losses in photodetection and in quantum-state measurements},
  author={Kiss, T and Herzog, U and Leonhardt, Ulf},
  journal={Phys. Rev. A},
  volume={52},
  number={3},
  pages={2433},
  year={1995},
  publisher={APS}
}

@article{krishnaswamy2024experimental,
  title={Experimental retrieval of photon statistics from click detection},
  author={Krishnaswamy, Suchitra and Schlue, F and Ares, Laura and Dyachuk, Vladyslav and Stefszky, Michael and Brecht, Benjamin and Silberhorn, Christine and Sperling, Jan},
  journal={Phys. Rev. A},
  volume={110},
  number={2},
  pages={023717},
  year={2024},
  publisher={APS}
}

@article{miatto2012spatial,
  title={Spatial Schmidt modes generated in parametric down-conversion},
  author={Miatto, Filippo M and di Lorenzo Pires, H and Barnett, Stephen M and van Exter, Martin P},
  journal={Eur. Phys. J. D },
  volume={66},
  number={10},
  pages={263},
  year={2012},
  publisher={Springer}
}

@article{eberly2006schmidt,
  title={Schmidt analysis of pure-state entanglement},
  author={Eberly, JH},
  journal={Laser physics},
  volume={16},
  number={6},
  pages={921--926},
  year={2006},
  publisher={Springer}
}

@article{krapick2014bright,
  title={Bright integrated photon-pair source for practical passive decoy-state quantum key distribution},
  author={Krapick, Stephan and Stefszky, MS and Jachura, Michal and Brecht, Benjamin and Avenhaus, Malte and Silberhorn, Christine},
  journal={Phys. Rev. A},
  volume={89},
  number={1},
  pages={012329},
  year={2014},
  publisher={APS}
}

@article{klyshko1977utilization,
  title={Utilization of vacuum fluctuations as an optical brightness standard},
  author={Klyshko, DN},
  journal={Sov. J. Quantum Electron.},
  volume={7},
  number={5},
  pages={591},
  year={1977},
  publisher={IOP Publishing}
}

@article{sperling2014quantum,
  title={Quantum state engineering by click counting},
  author={Sperling, J and Vogel, W and Agarwal, GS},
  journal={Phys. Rev. A},
  volume={89},
  number={4},
  pages={043829},
  year={2014},
  publisher={APS}
}

@misc{multimodePDC,
  title={ \textrm{Following the treatment in Ref.~\cite{borrero2025advancing} one can compute the values of $g_{\text{h}}^{(2)}$ in the multimode case by replacing the thermal photon-number distribution of the single-mode PDC emission with a Poissonian one denoted by $P_{n}= \bar{n}^ne^{-\bar{n}}/n!$ with the mean-photon number of $\bar{n}$. Additionally, one needs to take into account that in the multimode case CAR $ = \frac{\sum_{n=0}^{\infty}n^2P_{n}(\bar{n})}{\left(\sum_{n=0}^{\infty}nP_{n}(\bar{n})\right)^{2}} = 1 + \frac{1}{\bar{n}}$ }}
  }

\end{document}